\pdfoutput=1
\documentclass[bibnotes,prb,amsmath,citeautoscript,twocolumn,%preprint,%
               preprintnumbers,amssymb,amsfonts,showpacs,
               letterpaper]{revtex4}
\usepackage{graphicx} % to include images
\begin{document}

% -----------------------------------------------------------------------------

\title{Structural effects on the spin-state transition in epitaxially strained LaCoO$_3$ films}
  \author{James M.\ Rondinelli}
  \affiliation{Materials Department, University of California, Santa Barbara, 
	       CA, 93106-5050, USA}
  \author{Nicola A.\ Spaldin}
     \email[Address correspondence to: ]{nicola@mrl.ucsb.edu}
  \affiliation{Materials Department, University of California, Santa Barbara, 
	       CA, 93106-5050, USA}
\date{\today}
% -----------------------------------------------------------------------------

% -----------------------------------------------------------------------------
\begin{abstract}
Using density functional theory within the LSDA + $U$ method, we investigate the
effect of strain on the spin state and magnetic ordering in perovskite 
lanthanum cobaltite, LaCoO$_3$.
We show that, while strain-induced changes in lattice parameters are insufficient
to stabilize a non-zero spin state, additional heteroepitaxial symmetry constraints -- in particular the 
suppression of octahedral rotations -- stabilize a ferromagnetic intermediate-spin state.
By comparing with experimental data for the bulk material, we calculate  
an upper bound on the Hubbard $U$ value, and describe the role 
that the on-site Coulomb interaction plays in determining the spin-state configuration.
\end{abstract}
% -----------------------------------------------------------------------------

% -----------------------------------------------------------------------------
% -----------------------------------------------------------------------------
\maketitle
% -----------------------------------------------------------------------------

% -----------------------------------------------------------------------------

\section{Introduction}
The desire to control magnetism with external perturbations other than magnetic fields 
has motivated much recent research on the strain- and electric field-
response of magnetic materials
\cite{Ramesh/Spaldin:2007,Brinkman_et_al:2007,Yamada_et_al:2004}.
Such control might enable future {\it Mottronic} applications, in which
small external perturbations drive transitions between competing electronic,
orbital, charge and spin orderings, causing drastic changes in properties.
Thin film heterostructures containing transition metal oxides 
are proving particularly promising in this emerging field; for example, electric-field
switching of magnetization has been achieved through exchange-bias coupling 
of ferromagnetic Co$_{0.9}$Fe$_{0.1}$ to multiferroic BiFeO$_3$ \cite{Chu_et_al:2008}, and 
substrate induced strain has been used to tune magnetic 
interactions\cite{Yamada/Tokura:2006} in magnetoelastic composites (see 
Ref.\ ~\onlinecite{Eerenstein/Mathur/Scott:2006} for a review).
Recent reports\cite{Fuchs_et_al:2007,Fuchs_et_al:2008,Herklotz_et_al:2008,Freeland/Ma/Shi:2008} of a substrate-dependent spin state in epitaxial films of 
LaCoO$_3$ 
are of particular interest since they suggest a route to switching magnetism
off (low spin diamagnetic $d^6$ Co$^{3+}$) and on (intermediate- 
or high-spin Co$^{3+}$). 
Lanthanum cobaltite is a rhombohedral ($R\bar{3}c$) perovskite that has been of continued interest for 
the last half-century, due in part to the many magnetic phase transitions that occur as 
a function of temperature, pressure and chemical doping \cite{Goodenough:1958,
Saitoh_Fujimori:1997, Medarde_et_al:2006,Vanko_et_al:2006,Knizek_Novak:2006}.
These transitions are a consequence of the competing crystal-field splitting energy 
($\Delta_{\rm CF}$), Hund's exchange energy ($J_{\rm H}$)  and $d$-orbital valence bandwidth ($W$), 
which are similar in magnitude, resulting in  low-, intermediate- or high-spin  
$d^6$ Co$^{3+}$, depending on the details of the system.
In the ground state (T = 0~K), LaCoO$_3$ is a diamagnetic insulator with a low-spin 
($S$=0, $t_{2g}^6e_g^0$) Co$^{3+}$ configuration. It is thermally excited to a
paramagnetic
intermediate- ($S$=1, $t_{2g}^5e_g^1$)  or high-spin ($S$=2, $t_{2g}^4e_g^2$) 
semiconducting state above approximately 95~K \cite{Radaelli_Cheong:2002}.
The nature of this spin-state transition is still under debate: 
inelastic neutron scattering \cite{Podlesnyak_et_al:2006}, x-ray 
absorption spectroscopy (XAS) and magnetic circular dichroism experiments 
\cite{Haverkort_et_al:2006} suggest a first-order transition to the high-spin 
(HS) state, while other x-ray photoemission (XPS) and XAS spectra in 
addition to electron energy loss (EELS) spectroscopy suggest the 
intermediate-spin (IS) state \cite{Abbate_Fujimori:1993,Masuda_Kato:1993,
Saitoh_Fujimori:1997,Klie_Leighton:2007}.
Similarly, Hartree-Fock cluster \cite{Zhuang:1998} and full-potential DFT 
calculations \cite{Knizek_Novak:2006} suggest the HS 
state is more stable than the IS state, while other LSDA + $U$ calculations 
obtain the reverse result \cite{Korotin_Khomskii:1996,Pandey_et_al:2008}.
In contrast to the bulk behavior, recent studies on LaCoO$_3$ thin films report
ferromagnetism with a field-cooled magnetization of 0.37 $\mu_B$/Co ion
on a substrate that causes 1.84\% tensile strain 
\cite{Fuchs_et_al:2007,Fuchs_et_al:2008}. 
Although ferromagnetic hysteresis loops have been recorded by several groups,\cite{Fuchs_et_al:2007,Fuchs_et_al:2008,Herklotz_et_al:2008,Freeland/Ma/Shi:2008,Pinta/Fuchs_et_al:2008} 
it remains unclear experimentally whether such magnetism is an intrinsic feature 
of strained LaCoO$_3$, or whether it arises from sample off-stoichiometry 
(ferromagnetism induced by hole doping is observed in bulk Sr-rich LaCoO$_3$ 
samples \cite{S-Rodriguez/Goodenough:1995}), or is a surface effect resulting
from the change in coordination of surface Co ions (also recently reported
in bulk samples \cite{Yan/Zhou/Goodenough:2004}).
In this work, we use {\it ab-initio} calculations based on density functional
theory (DFT) to show that epitaxial 
strain can indeed drive a spin-state transition to a ferromagnetic state
in stoichiometeric LaCoO$_3$. The transition is not caused, however,
by strain-induced changes of the lattice constants, but rather relies
on interface-induced changes in the tilt pattern of the CoO$_6$ octahedra.
\section{Computational Details}
We use the the projector augmented plane wave (PAW) method of DFT \cite{Bloechl:1994}, 
as implemented in the Vienna \emph{Ab Initio} Simulation Package ({\sc vasp}) 
code \cite{Kresse/Furthmueller_PRB:1996,Kresse/Joubert:1999}.
To accurately describe the exchange and correlation, we use the spherically averaged 
form of the rotationally invariant local spin density approximation + Hubbard $U$ 
(LSDA+$U$) method \cite{Anisimov/Aryasetiawan/Liechtenstein:1997,Dudarev_et_al:1998} 
with one effective Hubbard parameter $U_{\rm eff} = U - J$, and treat the double 
counting term within the fully localized limit. 
We use 
the supplied {\sc vasp} PAW pseudopotentials ({\sc La\_s, Co, O\_s}) with the 
5$p^6$5$d^1$6$s^2$ valence configuration for La, 4$s^1$3$d^8$ for Co, and 2$s^2$2$p^4$ for O.
Other technical details include a plane wave energy cutoff of 550~eV, a $7\times7\times7$ 
$k$-point grid to sample the Brillouin zone, and the tetrahedron method with Bl{\"o}chl 
corrections \cite{Bloechl/Jepsen/Andersen:1994} and an $11\times11\times11$ $k$-point grid 
to calculate the densities of states.

\section{Bulk L\lowercase{a}C\lowercase{o}O$_3$}

\subsection{Correlation Effects}%

The LSDA$+$ Hubbard $U$ approach has been successful in treating static correlations 
in transition metal oxides; however the selection of an appropriate $U_{\rm eff}$ is rarely 
straightforward, and a number of methods exist for determining suitable values. 
These include experimental measurement from photoemission spectroscopy \cite{Fujimori_et_al:1992}, self-consistent calculations  \cite{Anisimov/Gunnarsson:1991,Cococcion/Gironcoli:2005,Madsen_Novak:2005} 
and educated guesswork.
LaCoO$_3$ represents a particularly difficult case because of the strong
dependence of orbital occupation  -- which affects the polarizability and
screening, and in turn the $U_{\rm eff}$ -- on pressure and strain.
Indeed, XPS experiments have found the $d$-$d$ Coulomb correlations to range 
from 3.5 to 7.5~eV \cite{Sarma_et_al:1995,Saitoh_Fujimori:1997} depending on 
the structural details of the samples. 
Previous single-site and two-site configuration-interaction cluster 
calculations have obtained values from 4 to 5.5~eV.\cite{Abbate_Fujimori:1993,Saitoh_Fujimori:1997}
On the other hand, recent first-principles calculations suggest that the spin state 
is independent of the choice $U$,\cite{Klie_Leighton:2007} and values as large  
as 9~eV have been used to study temperature dependent spin-state 
transitions.\cite{Korotin_Khomskii:1996}
Each method does agree however that the $d$-$d$ electron repulsion for the low- and 
intermediate-spin states is approximately the same.
Due to these discrepancies in the literature, in the first part of this 
study we revisit the effects of electron repulsion on the spin-state and 
orbital occupation in bulk LaCoO$_3$. 
\begin{figure}
\includegraphics[width=0.48\textwidth]{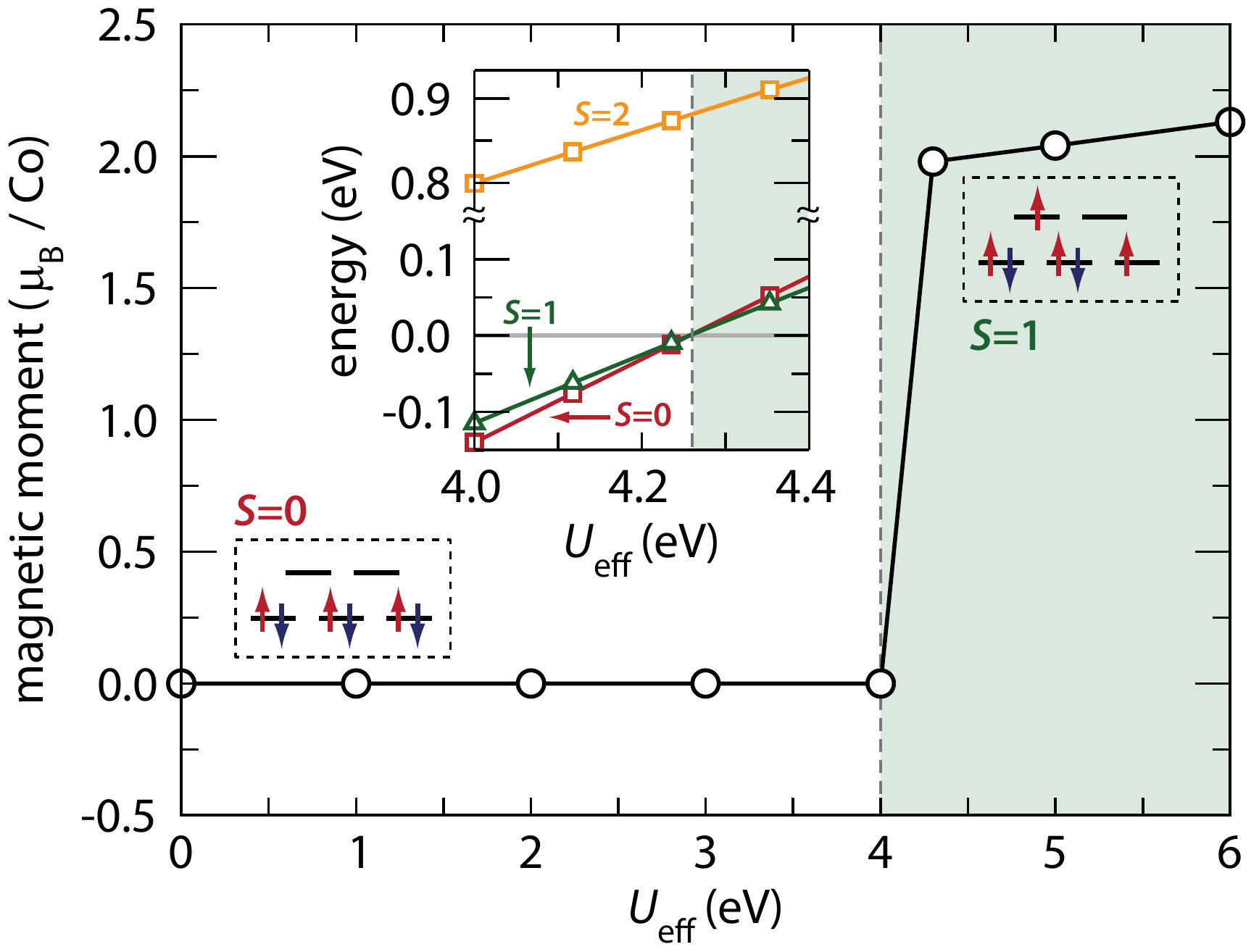}
\caption{\label{fig:critical_U}
(Color online) Calculated magnetic moment as a function of $U_{\rm eff}$ in bulk 
rhombohedral LaCoO$_3$ with the corresponding orbital energy diagrams. 
({\sc inset}) Relative energies per formula unit of the $S=0$, $S$=1 and $S$=2 
spin states.
}
\end{figure}
We begin by calculating the critical $U$ value that induces a spin 
state transition in the bulk rhombohedral structure by calculating the 
total energies for each spin state as a function of $U_{\rm eff}$.
Since the bulk zero kelvin state is known to be low spin, this requirement then 
provides a bound on allowable $U_{\rm eff}$ values for studying the 
experimentally observed ferromagnetic thin films.
Here we use the low temperature experimental structural parameters found in Ref.\ ~\onlinecite{Radaelli_Cheong:2002}
($a=5.275$~\AA, $\alpha = 61.01^\circ$) and fully relax the internal coordinates 
with $U_{\rm eff} = 0$~eV until the forces are less than 1~meV~\AA$^{-1}$.
Within the LSDA, we find the correct ground state structure: a diamagnetic insulator 
with a 0.45~eV band gap, which is close to the measured optical 
gap \cite{Sarma_et_al:1995}.
Our calculated energies and magnetic moments as a function of $U_{\rm eff}$ 
are shown in Figure \ref{fig:critical_U}. The most striking finding is
that the experimentally observed $S=0$ ground state is only stable for $U_{\rm eff}$
values less than 4.0 eV.
Therefore we regard 4.0 eV as an upper bound on $U_{\rm eff}$ for LaCoO$_3$.
In Figure~\ref{fig:critical_U}(inset) the relative energies of
the diamagnetic $S=0$, IS $S=1$, and HS $S=2$ states are also shown.
[For these comparison calculations we impose ferromagnetic (FM) order in 
the IS and HS Co sublattice so that we can fix the total spin moment.]
As expected, as $U_{\rm eff}$ is increased, spin pairing in the $t_{2g}$
manifold becomes less favorable as the energy gain from the Hund's exchange 
energy exceeds the energy cost in creating a singlet state and thereby reduces  
the relative energies of higher spin states.
Such correlation-induced spin-state transitions, in which higher $U_{\rm eff}$
values induce states with higher magnetic moments, have previously been found in 
a number of other transition metal compounds.
\cite{Lebegue/Pillet/Agnyan:2008,Werner/Millis:2007,Tsuchiya/de_Gironcoli:2006,Baettig/Ederer/Spaldin:2005} 
Co$^{3+}$ is a particularly interesting case, however, because the low-spin state is
non-magnetic and so the spin-state transition increases the magnetic moment to a finite
value from an initial value of zero. 
Interestingly, at the transition to the $S=1$ state, we find $U_{\rm eff}/W$=0.27,
which is low compared to most moderately- or strongly-correlated magnets;
in addition, $W$ is largely independent of $U_{\rm eff}$ (not shown).
For all $U_{\rm eff}$ values we find that the HS state is more than 
1.0~eV higher in energy than the IS or LS states.
The critical  $U_{\rm eff}$ value we have determined with this approach is in good agreement with x-ray photoemission experiments 
reported in Ref.\ ~\onlinecite{Sarma_et_al:1995} and recent first-principles 
calculations in Ref.\ ~\onlinecite{Pandey_et_al/2008}.
Therefore, for the remainder of this study, we strictly use values 
of $U_{\rm eff}<4.0$~eV unless noted otherwise.
The strong dependence of the ground-state spin configuration on $U_{\rm eff}$ 
partly explains the inconsistencies between different first-principles calculations in 
describing the evolution of the spin-state transition.

\subsection{Electronic Structure}
\begin{figure}
\includegraphics[width=0.48\textwidth]{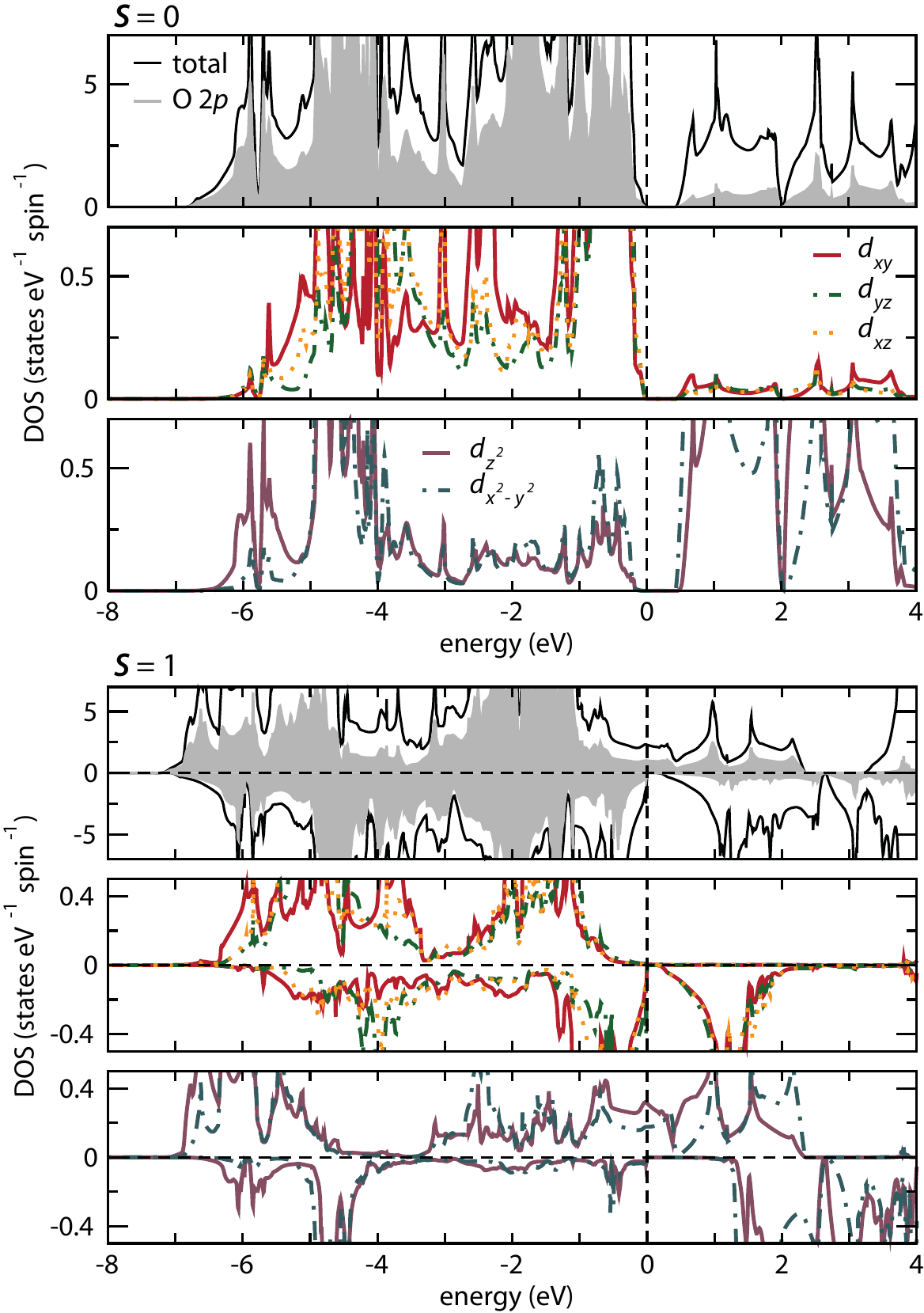}
\caption{\label{fig:is_dos}
(Color online) Spin- and orbital-resolved densities of states for $S=0$ 
({\sc upper}) and $S=1$ ({\sc lower}) rhombohedral LaCoO$_3$ ($U_{\rm eff}=3.0$~eV).
In the low-spin state, a diamagnetic insulator is found, while a 
half-metallic ground state is found with a local magnetic moment of 
1.8~$\mu_B$ per Co atom in the intermediate spin state.}
\end{figure}

Before investigating the effects of strain on the magnetic behavior we 
describe the nature of the unusual intermediate $S=1$ state of LaCoO$_3$ compared 
to the diamagnetic $S=0$ state. 
In the molecular cluster limit [Figure \ref{fig:critical_U}(inset)], when 
$\Delta_{\rm CF} > J_{\rm H}$ the low-spin 
configuration is favored, while when $\Delta_{\rm CF} < J_{\rm H}$ the high-spin state 
dominates due to the gain in exchange energy from the parallel alignment of spins; the
intermediate spin state might be expected when $\Delta_{\rm CF} \approx J_{\rm H}$. 
Furthermore, as evident in Figure \ref{fig:is_dos}, hybridization and covalency between the 
O $2p$ and Co $3d$ states causes dramatic deviations from the simple molecular cluster picture by causing
strong broadening of the bands; in particular the Co $e_g$ orbitals span more than 
11 eV in energy.
In Figure \ref{fig:is_dos} we show our calculated electronic densities of states  
for the $S=0$ and constrained $S=1$ ferromagnetically ordered LaCoO$_3$ at the experimental lattice parameter with a $U_{\rm eff}=3.0$~eV.
In the low-spin state, the Co $t_{2g}$ manifold is fully occupied with 
triply degenerate $d_{xy}$, $d_{xz}$ and $d_{yz}$ orbitals; the valence 
band is formed by a mixture of these Co states and O 2$p$ orbitals. 
The doubly degenerate $d_{x^2-y^2}$ and $d_{z^2}$ orbitals which 
make up the $e_g$ manifold form the conduction band.
In the intermediate-spin state, on the other hand, broadening of the band widths 
causes the majority up-spin $e_g$ band to extend below the Fermi energy, and the 
minority $t_{2g}$ band to extend above. 
The hole in the $t_{2g}$ manifold consists of a superposition of the minority 
$\frac{1}{\sqrt{3}}(d_{xy}+d_{yz}+d_{xz})$ orbitals, and the ``missing'' electron 
occupying the lower part of the majority $e_g$ band is in a 
$\frac{1}{\sqrt{2}}(d_{z^2}+d_{x^2-y^2})$ state.
This behavior causes our calculated ferromagnetic intermediate-spin state to be
half-metallic. Note, however, that bulk LaCoO$_3$ is in fact an insulator up to room 
temperature \cite{Tokura_Takagi:1998}; we will return to this discrepancy later.
\section{Strained L\lowercase{a}C\lowercase{o}O$_3$}
Spin-state transitions in bulk LaCoO$_3$ are known to occur as a 
function of unit cell volume\cite{Radaelli_Cheong:2002}, where the  
tendency for an excited spin state is favored for larger unit cell 
volumes due to competition between the energy penalty in forming a singlet 
state and the gain in Hund's exchange energy favoring ferromagnetic alignment of spins.
This is evidenced by the fact that high spin Co$^{3+}$ has a larger ionic 
radius (0.61~\AA), compared to the low spin Co$^{3+}$ (0.55~\AA).\cite{Koehler_Wollan:1956}
Indeed, this is consistent with our calculations for the ideal cubic pervoskite 
where the LSDA$+U$ equilibrium volume of the intermediate state is approximately 
2\% larger than that of the low spin configuration.
In this section we examine likely crystallographic structural distortions in thin films to determine whether 
they cause a spin-state crossover.
In general, heteroepitaxial strain from coherent growth on a substrate
with a mismatched lattice constant can modify the structure by changing the
lattice parameters, symmetry or chemistry at the interface.
Therefore we explore whether computations that incorporate changes in Co--O--Co 
bond angles, Co--O bond lengths, unit cell volume, or combinations of these effects, 
are able to reproduce the ferromagnetic state which is observed experimentally in 
thin film LaCoO$_3$.

\subsection{Effect of Changes in Lattice Parameter}
The low temperature rhombohedral structure of LaCoO$_3$ belongs to the 
($a^-a^-a^-$) Glazer tilt system, in which successive octahedra rotate in 
opposite senses along each crystallographic direction. 
The Co--O--Co bond angle is approximately 166$^\circ$.
The importance of such octahedral rotations on the electronic properties 
of thin film perovskite oxides has been the subject of many recent reports, 
particularly in the context of their effect on ferroelectricity 
\cite{Singh/Park:2008,Hatt/Spaldin:2007}.
An effect on {\it magnetic} properties is also likely, because changes in 
Co--O--Co bond angles can strongly affect the magnetic superexchange 
interactions.
In this section we investigate whether strain-induced changes in these bond angles 
are sufficient to stabilize the intermediate-spin state.
\begin{figure}
\includegraphics[width=0.48\textwidth]{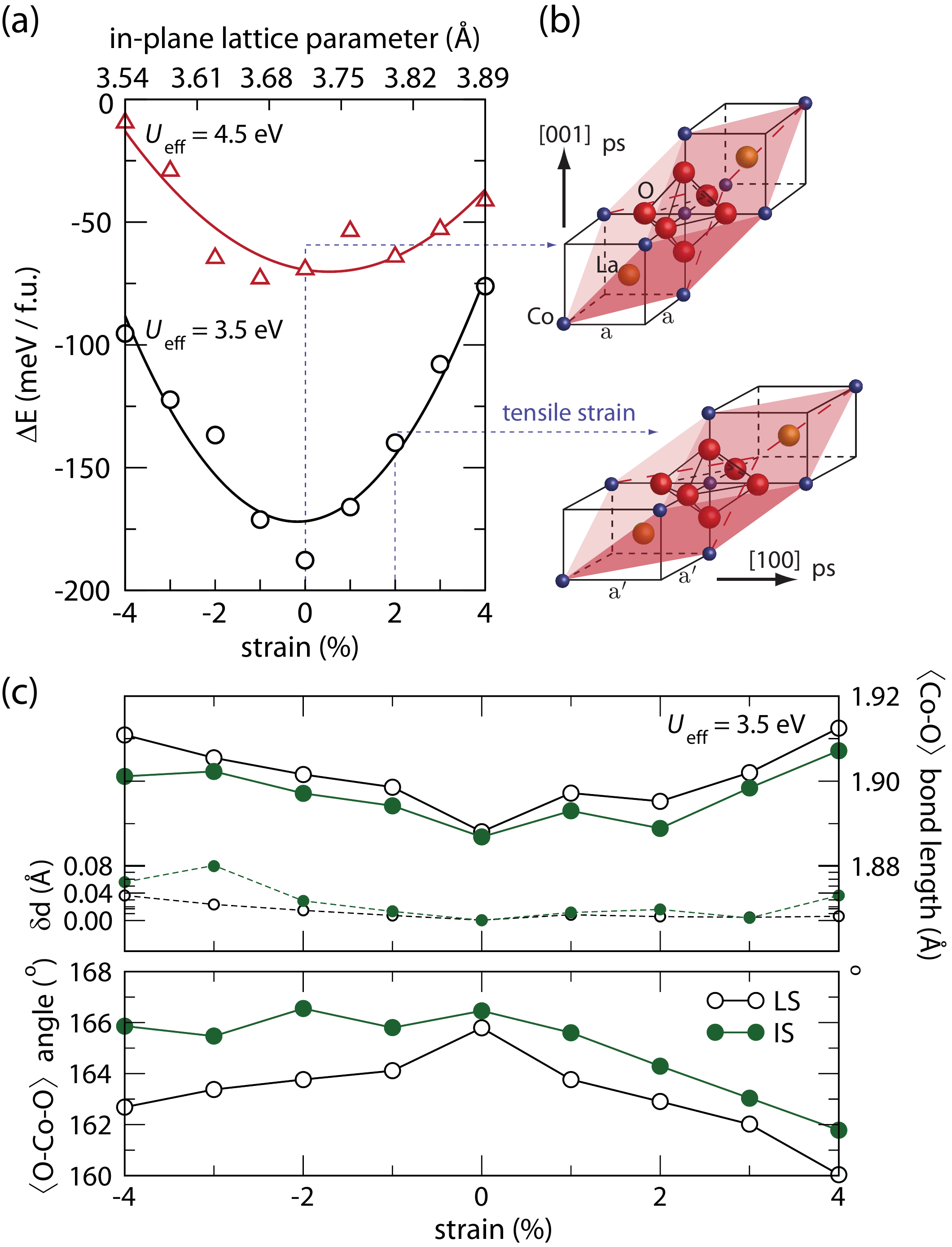}
\caption{\label{fig:rhom_strain}(Color online) (a) 
Energy difference between the diamagnetic low-spin and ferromagnetic intermediate-spin states, 
$\Delta E$, 
as a function of epitaxial strain applied in the pseudo-cubic (ps) (100) plane 
of the rhombohedral structure (b) shown relative to the cubic perovskite.
(c) The relaxed mean Co--O bond lengths, bond differences $\delta d$ between the long- and short 
bonds in the CoO$_6$ octahedra, and the  O--Co--O bond angles for
$U_{\rm eff} = 3.5 eV$. The lines are a guide for the eye.
}
\end{figure}
In Figure \ref{fig:rhom_strain}(a), we plot our calculated energy difference between low-
and intermediate-spin states with ferromagnetic order 
as a function of strain applied to the pseudo-cubic (100) 
plane with respect to the LSDA equilibrium volume. Values are shown for two $U_{\rm eff}$ 
values, 3.5 and 4.5 eV, chosen to be above and below the critical $U_{\rm eff}$ value of 
4.0 eV that we established in the previous section.
(The high spin state is not shown, since it is $\sim$1~eV higher in energy at 
all strain values.) 
At each in-plane strain value we adjust the out-of-plane lattice parameter 
and rhombohedral angle to maintain the bulk equilibrium volume, then fully
relax the atomic positions. 
We find that, for $U_{\rm eff}$ below our calculated critical value, the 
low-spin ground state is stable up to strains of 4\%; therefore we predict that 
strain-induced changes in lattice parameters alone are insufficient to cause a 
spin-state transition in LaCoO$_3$ up to reasonable strain values. 
Even our unphysically large $U_{\rm eff}$ of 4.5 eV does not induce a transition 
to the intermediate spin state until just beyond 4\% compressive strain.
The half-metallic state remains stable until strain values above 3\% ($U_{\rm eff}=3.5$~eV) 
and 1\% ($U_{\rm eff}=4.5$~eV), when the electronic structure becomes fully metallic. 
To understand the absence of spin-state transition with strain, 
we plot in Figure \ref{fig:rhom_strain}(c) the evolution of the mean 
Co--O bond length and the mean Co--O--Co bond angle as a function of 
strain for $U_{\rm eff} = 3.5$~eV.
With either compressive or tensile strain, the average Co--O bond length increases 
from the equilibrium (zero-strain) value. All bond lengths increase uniformly, however,
such that the CoO$_6$ octahedra remains perfectly octahedral, and the ideal $O_h$ 
crystal field is maintained.
This is supported by the bond length differences between the long- and short- Co--O bonds  
($\delta d$) in the CoO$_6$ octahedra.
We find significant changes in the Co-O-Co bond angles, particularly for tensile
strain. In many magnetic perovskites, such large changes in bond angles are sufficient
to change the magnetic ordering.\cite{Subramanian_et_al:1999,Retuerto/Attfield:2007}
We believe that the absence of spin crossover in
this case is due to the exceptionally broad bandwidth of the Co $e_g$ orbitals, which 
reduces the exchange energy gain from spin polarization. 

\subsection{Effect of Octahedral Rotations and Distortions}
Next we isolate the influence of these octahedral rotations by manually 
disabling them while applying strain.
Our motivation is two-fold. First, there is experimental evidence suggesting that 
LaCoO$_3$ grows in such a `cube-on-cube' manner on many substrates 
\cite{Fuchs_et_al:2007,Fuchs_et_al:2008,Suzuki_et_al:2008}. Second, disabling
the octahedral rotations causes the system to respond to strain by changing the
local symmetry around the Co ion; therefore we can examine the influence of the
crystal field on the spin-state transition.
The no-rotations constraint is imposed by using a five atom unit cell which prohibits 
rotations by symmetry; as a side-effect this also imposes ferromagnetic ordering.  
(We later examine if this is indeed the preferred magnetic ordering.)
We begin by setting the in-plane pseudo-cubic lattice parameter ($a$) to that of the 
experimental substrate (LaAlO$_3$)$_{0.3}$(Sr$_2$AlTaO$_6$)$_{0.7}$ (LSAT) value (3.87 \AA) with the optimized LSDA$+U$ volume ($c/a=0.955$), and relax the internal coordinates and out-of-plane ($c$) lattice 
constant; the resulting structure is diamagnetic and 270~meV higher in energy than 
the LSDA equilibrium $R\bar{3}c$ LS structure.
\begin{figure}
\includegraphics[width=0.45\textwidth]{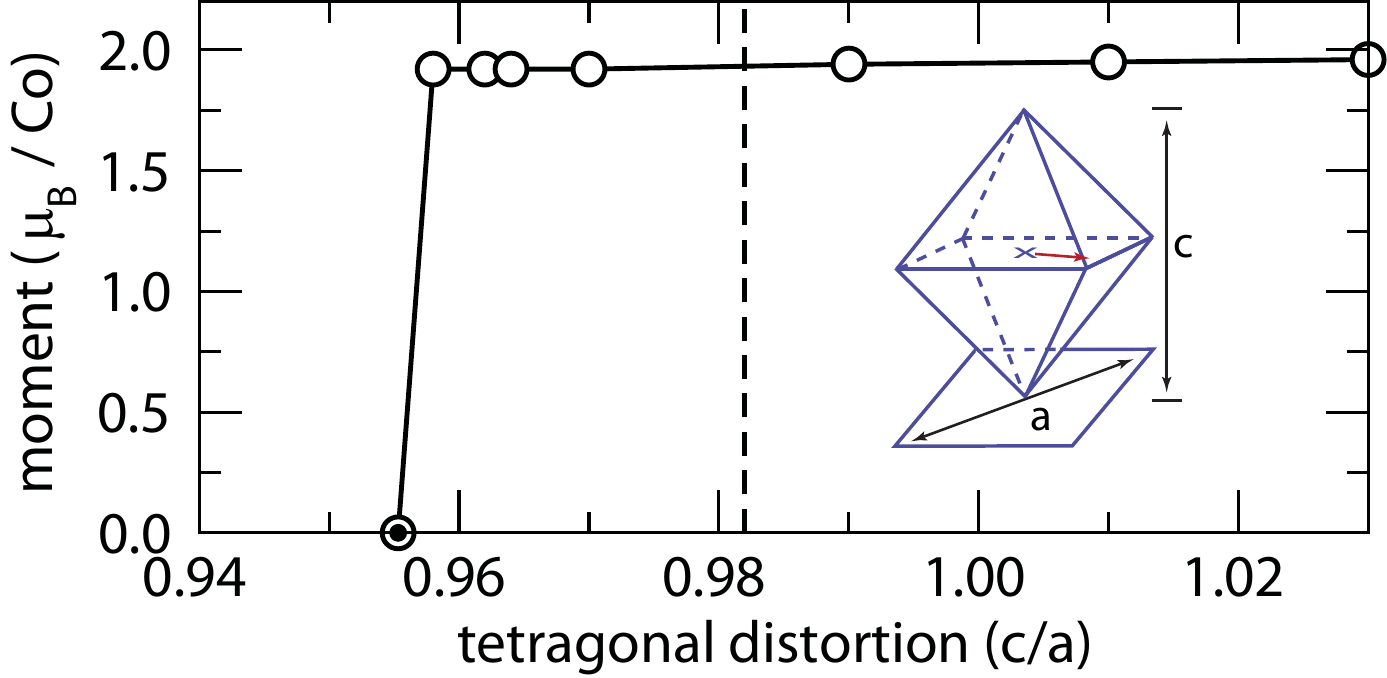}
\caption{\label{fig:tet_mag}(Color online)
Calculated magnetic moment as a function of $c/a$ for tetragonal LaCoO$_3$ 
with $a$ fixed to the experimental LSAT lattice parameter (3.87 \AA).
The dashed line (filled circle) indicates the experimental (LSDA equilibrium) 
$c/a$ ratio.
({\sc inset}) Schematic representation of Co displacement within the 
equatorial octahedral plane (arrow).}
\end{figure}
Next, we apply uniaxial strain by varying the $c/a$ ratio with 
$a$ fixed to the experimental LSAT value, and
show the resulting calculated magnetic moment in Figure \ref{fig:tet_mag}. 
Our main finding is that, at the experimental $c/a$ ratio, the 
IS state is lower in energy than the LS, and therefore we 
predict that the Co ions should be magnetic.
The origin of the stabilization of the intermediate state is the lifting
of the octahedral crystal field by the tetragonal symmetry adopted when the 
octahedral rotations are disabled.
Small ($< 1 \%$) uniaxial expansion of the out-of-plane lattice constant 
modifies the tetragonal crystal field splitting sufficiently to favor 
occupation of the $e_g$ manifold. 
In contrast, when octahedral rotations are allowed, strain is accommodated through
changes in the rotation angles rather than through modification of the local
bond lengths around the Co ions; the local crystal field splitting is therefore
largely unchanged and the diamagnetic state remains stable.
Next we investigate whether the ferromagnetic order imposed so far for
computational convenience is indeed the lowest energy magnetic ordering  
by comparing its energy with those of the A-type antiferromagnetic 
({\sc a-afm}), and G-type antiferromagnetic ({\sc g-afm})
orderings for this structure.
The total energies for each structure are shown in Table \ref{tab:mag_ordering}
for both the experimental and optimized LSDA$+U$ $c/a$ values. 
The energies are given relative to the FM single unit cell configuration.
\begin{table}
\begin{ruledtabular}
\begin{tabular}{lcccc}%%{lcclcl} here
& $c/a$           &  {\sc fm}     & {\sc a-afm}   & {\sc g-afm}   \\
\hline
LSDA+$U$ &0.955&       0.0  (0.0)   &       n/a (0.0)\footnote{We were unable to converge
the A-type AFM arrangement.}&       +147   (2.8) \\
Exp.\ &0.982   &       0.0 (1.9)     &       +375 (2.7)   &       +180  (2.8)  \\
\end{tabular}
\end{ruledtabular}
\caption{\label{tab:mag_ordering}Total energy differences (in meV) for
various magnetic orderings within the tetragonal crystal structure relative to 
the five atom ferromagnetic unit cell. Values are given at the experimental
and optimized LSDA$+U$ $c/a$ ratios.
Calculated magnetic moments per Co atom in $\mu_B$ are given in parentheses.}
%\end{center}
\end{table}
We find that the {\sc g-afm} and {\sc a-afm} structures are 180 and 375 meV per formula 
unit higher in energy than the ferromagnetically ordered intermediate spin 
state at the experimental $c/a$ ratio. This is consistent with the experimentally observed 
ferromagnetism.
Intriguingly, the structure with disabled octahedral rotations relaxes to a 
polar space group ($C2/c$) with the Co$^{3+}$ ion moving 0.06~{\AA} off-center in the 
$ab$ plane.
By summing the formal ionic charges multiplied by their displacements from
their centrosymmetric positions we find an in-plane polarization of 17.5~$\mu$C/cm$^2$ at 
the experimental $c/a=0.982$ ratio.
Note that since our overall electronic ground state is metallic, we are unable to evaluate the electronic contribution to the polarization using the standard 
Berry's phase approach.
Indeed the onset of ferroelectric polarization caused by the disabling of
octahedral rotations in perovskite oxides has been noted in a number of calculations,
and is believed to result from the off-centering of ions to maintain a favorable
bond order \cite{Singh/Park:2008,Hatt/Spaldin:2007}.
However, since our electronic structure is overall metallic, we cannot predict 
ferroelectric behavior.

\section{Discussion}
\begin{figure}
\includegraphics[width=0.48\textwidth]{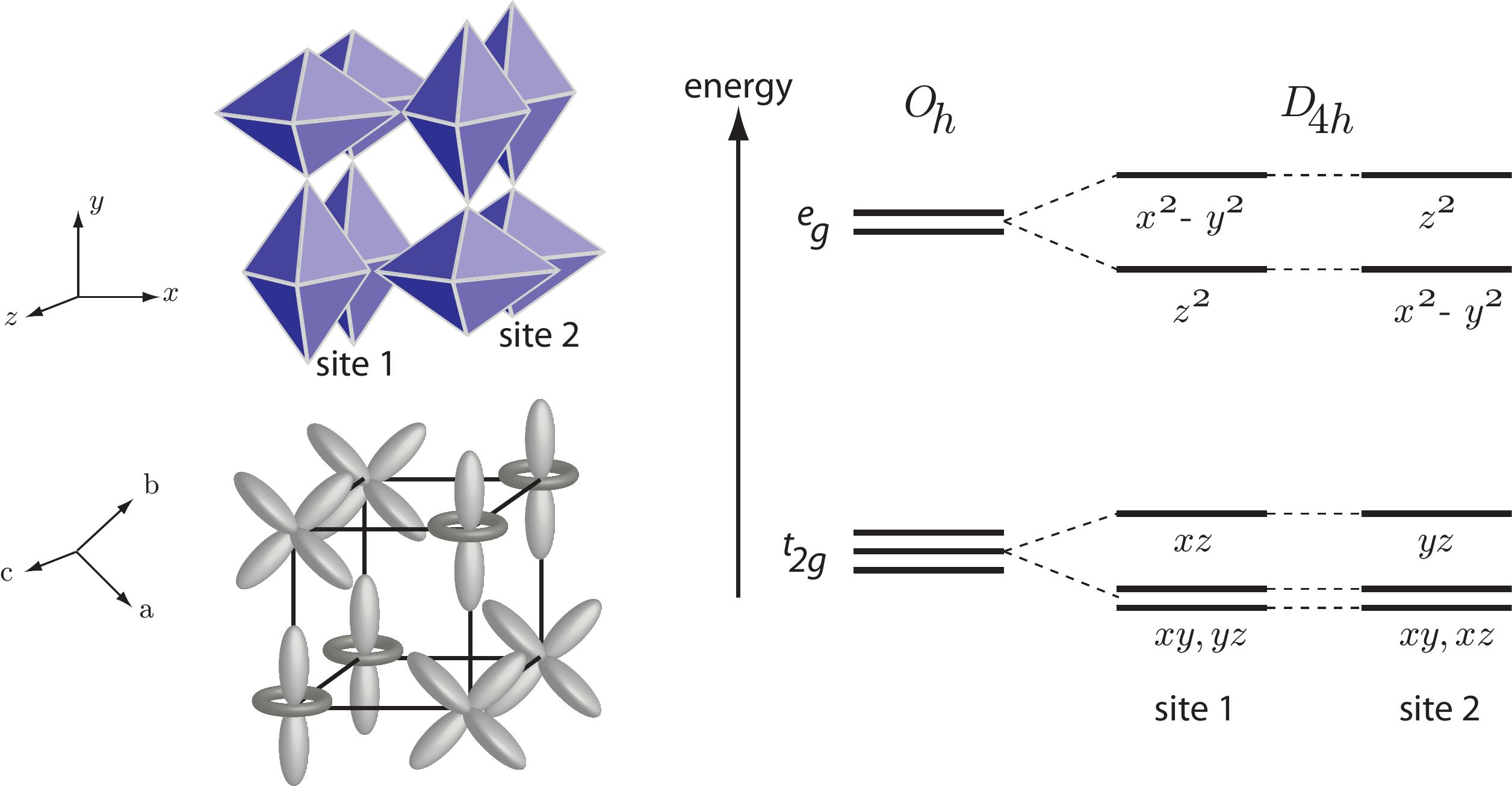}
\caption{\label{fig:d-JT}
The {\sc type-d} Jahn-Teller distorted structure is shown with the
possible $e_g$ orbital ordering configuration. The orbital degeneracy is
also split at each of the two sites, according to the orientation of the
elongation of the oxygen octahedra with respect to the $c$ lattice
parameter.}
\end{figure}

While our finding of ferromagnetism is consistent with recent experimental
reports, there are some important differences between our computations and
the experimental observations. First, and analogous to the bulk 
intermediate-spin case, 
our calculated tetragonal structure is
half-metallic, with a broad majority spin O $2p$ - Co $e_g$ band crossing
the Fermi level; experimentally the ferromagnetic films are found to be
insulating. 
In addition, the sizes of most measured magnetic moments are an 
order of magnitude smaller than our calculated value. Recent  
x-ray magnetic circular dichroism (XMCD) 
experiments on thin films, however, find a local Co moment of 1.2~$\mu_B$, which is in 
better agreement with our calculations \cite{Freeland/Ma/Shi:2008}. 
In this last section we attempt to reconcile our finding of a strain-induced half-metallic 
ferromagnetic arrangement with experimental reports of insulating LaCoO$_3$
on LSAT.
In particular, we explore likely Jahn-Teller distortions and orbital orderings
which are known to allow both ferromagnetism and insulating behavior in oxides
\cite{Mizokawa/Khomskii/Sawatsky:1999}.
We also examine the effect of including spin-orbit interactions in our 
calculations, since these couplings can make significant contributions to 
determining the orbital occupation in many transition metal oxides.
The possibility of an orbitally ordered state in LaCoO$_3$ was suggested previously 
in Ref.\ ~\onlinecite{Korotin_Khomskii:1996}, and unrestricted Hartree-Fock calculations 
on similar materials \cite{Mizokawa/Fujimori:1995} found that
small Jahn-Teller structural distortions can stabilize an insulating state.
Cooperative Jahn-Teller distortions (ranging from 1 to 6\% from low to room 
temperature) have indeed been demonstrated in LaCoO$_3$ with various techniques including
high-resolution x-ray diffraction \cite{Maris_Palstra_et_al:2003}, Raman 
scattering \cite{Ishikawa/Nohara/Sugai:2004}, and neutron 
diffraction \cite{Kozlenko/Glazkov_et_al:2007,Phelan/Yu/Louca:2008}.
Ref.\ ~\onlinecite{Maris_Palstra_et_al:2003} obtained
a monoclinic structure with $I2/a$  symmetry, consistent with the
{\sc a-afm} ordering seen in LaMnO$_3$.
Although the {\sc type-a} monoclinic structure was investigated previously in 
Ref.\ ~\onlinecite{Korotin_Khomskii:1996}, we revisit the possible orbital ordering 
available in strained LaCoO$_3$ by imposing from 1-6\% Jahn-Teller structural distortions of the CoO$_6$.
In the same way, we study {\sc type-d} ordering (space group $P4/mbm$), 
which has uniform orbital occupation along the 
$c$-direction and alternating orbital occupation in the $ab$-plane,
consistent with overall ferromagnetic superexchange (Figure \ref{fig:d-JT}).
This 2D antiferrodistortive behavior allows for $d_{x^2-y^2}$ and $d_{z^2}$ 
orbital ordering due to the splitting of the $e_g$ degeneracy from the structural 
distortions of the oxygen octahedra.
Similarly the threefold degenerate $t_{2g}$ orbitals also split into a 
two-fold degeneracy, that is lower in energy than the single non-degenerate state.
In all cases examined no $e_g$ orbital order is observed, in spite of the imposed octahedral 
structural distortions.
Due to the degeneracy of the Co $e_g$ states at the 
Fermi level, although the density of states is reduced, a half-metallic ferromagnetic 
state persists.
In contrast, the higher energy {\sc g-afm} structure considered earlier is insulating.

Finally, we examine the effect of spin-orbit interactions in LaCoO$_3$, since 
such coupling of the magnetic spin degrees of freedom to the lattice can 
change the relative level splitting and degeneracy of spin and orbital ground states 
in the 3$d$ transition metal oxides.\cite{Wu/Khomskii_et_al:2007,Maitra/Valenti:2007}
For the diamagnetic $S=0$ configuration we find that spin-orbit interaction has a 
negligible effect on the electronic structure as expected for a filled 
$t_{2g}$ manifold.
On the other hand, in a recent theoretical study\cite{Pandey_et_al/2008} it was suggested the $S=1$ state does show significant changes in the electronic structure near the Fermi level when spin orbit interactions are included.
In contrast to those results which used a generalized-gradient approximation (GGA$+U$) 
for the exchange-correlation functional, we do not find significant deviations in the 
electronic structure when spin-orbit coupling is included in the calculations.
This result is likely due to the difference in the description of our starting 
intermediate spin-state configurations: we begin with a half-metallic ground state, 
whereas in Ref.\ ~\onlinecite{Pandey_et_al/2008}, a fully metallic state is found with 
a high density of states at the Fermi level.
When spin-orbit coupling is included in the calculation, the large peak feature 
is naturally split into a doublet.
The origin of the two inconsistencies -- our half-metallic rather than insulating 
ground state, and larger magnetic moment per Co ion compared with experiment
-- might lie in the difficulties associated
with producing and characterizing high quality, uniform transition metal 
oxide films, or from a
failure of the LSDA+$U$ method to fully describe the complex orbital physics.
Future theoretical investigations should consider more sophisticated methods 
such as dynamical mean field theory, 
in which dynamical correlations (spin fluctuations) can in principle be treated explicitly.
Indeed, recent inelastic neutron scattering measurements report 
a small Jahn-Teller distortion which has short-range dynamical character 
\cite{Kozlenko/Glazkov_et_al:2007,Phelan_et_al:2006}.
This dynamic Jahn-Teller effect  is consistent with a proposed vibronic $e^1$--O--$e^0$ 
superexchange between intermediate-spin Co atoms \cite{Yan/Zhou/Goodenough:2004}, and 
would allow for fluctuations of AFM exchange which should reduce the magnetic moment.
On the experimental front, our calculations suggest that more detailed
characterization of the {\it local} electronic and structural properties
will be invaluable in understanding and exploiting the spin behavior of
LaCoO$_3$ films.

\section{Conclusions}
In summary, by comparing our calculated LSDA$+U$ spin state of bulk
LaCoO$_3$ with the measured low temperature behavior, we have determined 
a critical upper bound of 4 eV on the Hubbard $U$ parameter for LSDA$+U$ 
calculations for LaCoO$_3$.
Using our critical $U$ value, we have established that strain-induced changes in 
lattice parameters are insufficient to cause transitions to finite
magnetic moment spin states at reasonable values of strain.
Instead, if the cooperative octahedral tiltings and rotations are deactivated, 
intermediate-spin local moments are stabilized on the Co ions at small strain
values, and these order ferromagnetically.
Our results suggest a possible route to dynamically controlling magnetism using an
electric field, in superlattices of LaCoO$_3$ with a piezoelectric material.

\begin{acknowledgments}
We thank A.J.\ Hatt, R.\ Seshadri, T.\ Saha-Dasgupta, 
Y.\ Suzuki and J.W.\ Freeland for valuable discussions.
This work was supported by the NSF under the grant NIRT 0609377 (NAS) and 
a NDSEG fellowship sponsored by the DoD (JMR).
Portions of this work made use of MRL Central Facilities supported 
by the MRSEC Program of the National Science Foundation (DMR05-20415), 
the CNSI Computer Facilities at UC Santa Barbara (CHE-0321368), and 
the National Center for Supercomputing Applications under 
grant no.\ TG-DMR-050002S and utilized the SGI Altix {\sc cobalt} system.

\end{acknowledgments}

% Bibtex file here
%\bibliographystyle{apsrev}
%\bibliography{lco}

% Phew...EOF
\end{document}